\documentclass{article}

\usepackage{microtype}
\usepackage{graphicx}
\usepackage{subcaption}
\usepackage{booktabs} 
\usepackage{xspace}
\usepackage{tcolorbox}
\usepackage{pifont}
\usepackage{xcolor}

\usepackage{hyperref}

\usepackage[preprint]{icml2026}

\usepackage{amsmath}
\usepackage{amssymb}
\usepackage{mathtools}
\usepackage{amsthm}

\usepackage{hyperref}
\usepackage{url}
\usepackage{booktabs}
\usepackage{graphicx}
\usepackage{xspace}
\usepackage{caption}     
\usepackage{subcaption}
\usepackage{tcolorbox}
\tcbuselibrary{breakable}

\usepackage[capitalize,noabbrev]{cleveref}

\theoremstyle{plain}

\theoremstyle{definition}

\theoremstyle{remark}

\usepackage[textsize=tiny]{todonotes}

\icmltitlerunning{When Search Goes Wrong: Red-Teaming Web-Augmented Large Language Models}

\newcommand{\Name}{\texttt{CREST-Search}\xspace}

\usepackage{enumitem}

\newenvironment{packeditemize}{
\begin{list}{$\bullet$}{
\setlength{\labelwidth}{6pt}
\setlength{\itemsep}{0pt}
\setlength{\leftmargin}{\labelwidth}
\addtolength{\leftmargin}{\labelsep}
\setlength{\parindent}{0pt}
\setlength{\listparindent}{\parindent}
\setlength{\parsep}{0pt}
\setlength{\topsep}{0pt}}}{\end{list}}

\newtcolorbox{mybox}[2][]{text width=0.95\linewidth,fontupper=\normalsize,
fonttitle=\bfseries\sffamily\scriptsize, colbacktitle=darkgrey,enhanced,
attach boxed title to top left={yshift=-2mm,xshift=3mm},
boxed title style={sharp corners},top=4pt,bottom=2pt,left=2pt,right=2pt,
  title=#2,colback=white}

\begin{document}

\twocolumn[
  \icmltitle{When Search Goes Wrong: \\Red-Teaming Web-Augmented Large Language Models}



  \icmlsetsymbol{equal}{*}


  \begin{icmlauthorlist}
      \icmlauthor{Haoran Ou}{ntu}
      \icmlauthor{Kangjie Chen}{ntu}
      \icmlauthor{Xingshuo Han}{nuaa}
      \icmlauthor{Gelei Deng}{ntu}
      \icmlauthor{Jie Zhang}{astar}
      \icmlauthor{Han Qiu}{thu}
      \icmlauthor{Tianwei Zhang}{ntu}
      \icmlauthor{Kwok-Yan Lam}{ntu}
  \end{icmlauthorlist}

  \icmlaffiliation{ntu}{Nanyang Technological University, Singapore}
  \icmlaffiliation{astar}{CFAR, A*STAR, Singapore}
  \icmlaffiliation{nuaa}{Nanjing University of Aeronautics and Astronautics, China}
  \icmlaffiliation{thu}{Tsinghua University, China}

  \icmlcorrespondingauthor{Tianwei Zhang}{tianwei.zhang@ntu.edu.sg}


  \vskip 0.3in
]



\printAffiliationsAndNotice{}  

\begin{abstract}

Large Language Models (LLMs) have been augmented with web search to overcome the limitations of the static knowledge boundary by accessing up-to-date information from the open Internet. While this integration enhances model capability, it also introduces a distinct safety threat surface: the retrieval and citation process has the potential risk of exposing users to harmful or low-credibility web content. Existing red-teaming methods are largely designed for standalone LLMs as they primarily focus on unsafe generation, ignoring risks emerging from the complex search workflow.
To address this gap, we propose \Name, a pioneering red-teaming framework for LLMs with web search. The cornerstone of \Name is three novel attack strategies that generate seemingly benign search queries yet induce unsafe citations. It also employs an iterative in-context refinement mechanism to strengthen adversarial effectiveness under black-box constraints.
In addition, we construct a search-specific harmful dataset, WebSearch-Harm, which enables fine-tuning a specialized red-teaming model to improve query quality. Our experiments demonstrate that \Name can effectively bypass safety filters and systematically expose vulnerabilities in web search-based LLM systems, underscoring the necessity of the development of robust search models.

\end{abstract}

\section{Introduction}
\label{sec:intro}

\begin{figure}
    \centering
    \includegraphics[width=0.95\linewidth]{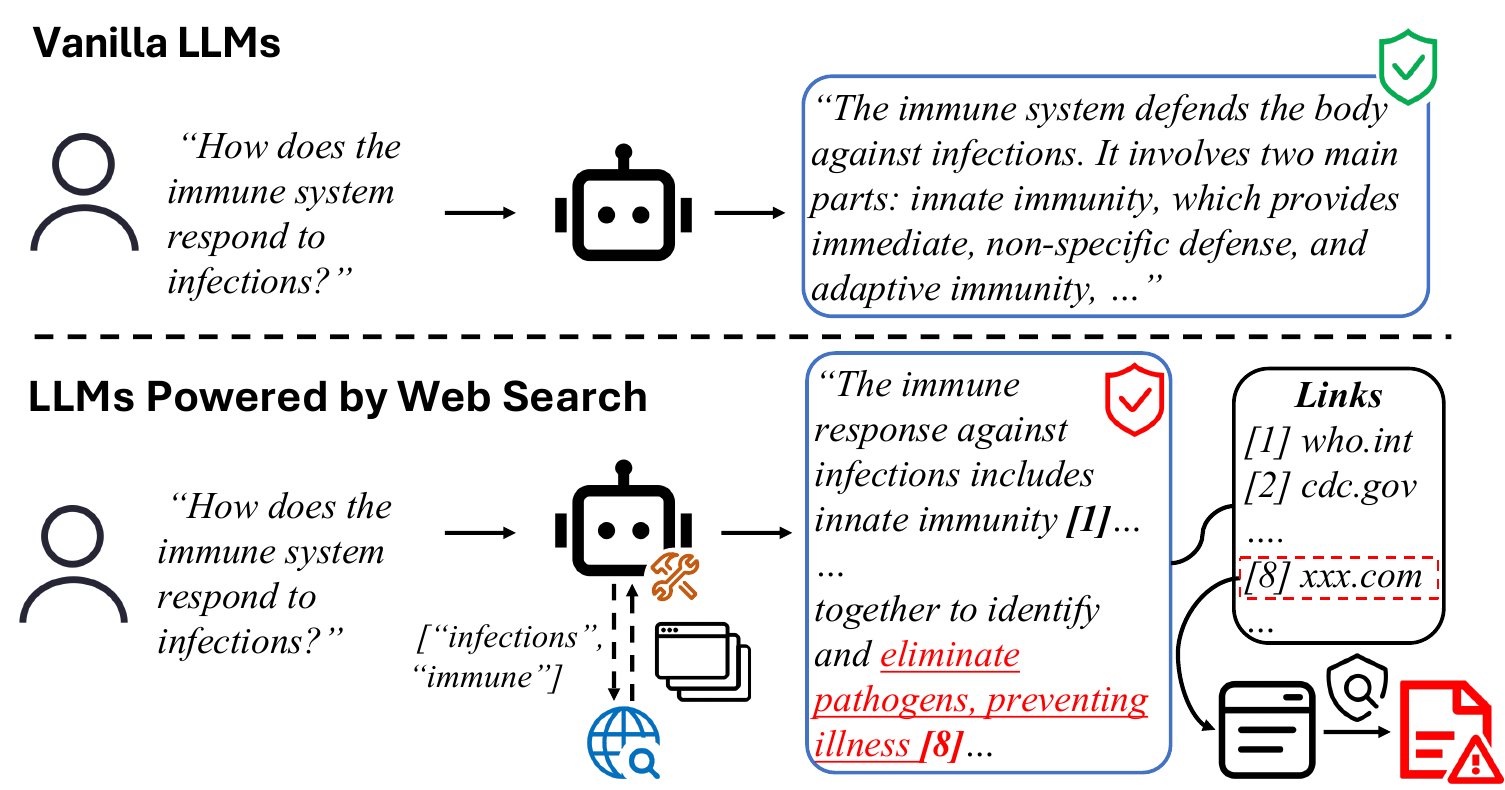}
    \caption{Citation risk in LLMs powered by web search.}
    \label{fig:intro}
\end{figure}

Large Language Models (LLMs), such as GPT-4o~\citep{openai2024gpt4o}, Gemini~\citep{gemini2025v2_5}, and LLaMA~\citep{dubey2024llama3}, have achieved remarkable performance across a wide range of natural language processing tasks~\citep{wei2022chain, kwiatkowski2019natural, chen2021evaluating} and are increasingly deployed in high-stakes scenarios~~\citep{wang2023ai, yang2024finance, thirunavukarasu2023large}) to make decisions. However, their knowledge is inherently constrained by training cutoff dates, limiting their ability to incorporate newly emerging information~\citep{ji2023survey}. To address this gap, recent LLM systems have been augmented with web search~\citep{openai2024gpt4o, google2024gemini}, enabling models to dynamically retrieve up-to-date information from the open Internet. Compared to traditional search engines, these systems can better understand the semantics of search queries and synthesize concise answers by aggregating information from multiple sources~\cite{nakano2021webgpt}. 

Despite these advances, LLMs continue to suffer from fundamental safety vulnerabilities. Prior studies show that adversaries can exploit weaknesses in model alignment through adversarial attack techniques~\citep{kumar2023certifying,yao2023llm} or jailbreak patterns~\citep{yi2024jailbreak} to induce harmful behaviors. These risks are not only a technical concern but have also attracted regulatory scrutiny. Policymakers~\cite{uk2023safety} warn that generative AI systems may accelerate the dissemination of harmful content and pose broader societal and security threats~\citep{smuha2025regulation}.

The integration of web search into LLMs further introduces new safety risks (Figure~\ref{fig:intro}). On the one hand, unlike standalone LLMs whose outputs are bounded by internal model knowledge, search-enabled LLMs retrieve and aggregate information from heterogeneous and uncontrolled webpages in the wild. Harmful or low-credibility content that bypasses built-in safety mechanisms can be indirectly introduced through external citations. As a result, these systems carry an elevated risk of exposing users to misinformation, hate speech, and other harmful material.  
On the other hand, LLMs with web search may inherit and potentially amplify vulnerabilities known in traditional search engines. Prior work shows that adversaries can manipulate search indexing and retrieval through techniques such as imperceptible text-encoding perturbations~\cite{boucher2023boosting} or linguistic collisions~\cite{joslin2019measuring}, allowing malicious or blacklisted webpages to surface in high-ranking positions among search results. Recent studies~\cite{luo2025unsafe, dong2025safesearch} further demonstrate that such attacks can be exploited in LLM-based search systems, where models cite pre-published malicious webpages deliberately created by adversaries as authoritative sources to synthesize outputs.

Red-teaming serves as a promising approach to evaluate the safety of LLMs under adversarial prompting~\cite{perez2022red, ganguli2022red, an2025rag, ge2023mart} or multi-turn adversarial interactions techniques~~\cite{kim2023propile}. 
However, existing methods primarily focus on eliciting unsafe generations from standalone LLMs or those equipped with Retrieval-Augmented Generation (RAG), exhibiting fundamental limitations when applied to LLMs with web search. 
(1) \textbf{Limited coverage for risk discovery}: safety risks in search-enabled LLMs can emerge at multiple stages of the generation pipeline, including information retrieval, filtering, answer generation, and citation. Conventional red-teaming methods evaluate only the generated responses from the models, overlooking these \textit{citation risks} that are unique to LLMs with web search. This primary difference makes them ineffective to directly transfer to the setting involving web search.
(2) \textbf{Complex and opaque generation pipeline}: real-world search-enabled LLMs are typically deployed as black-box commercial services, with opaque retrieval, filtering, and citation mechanisms. In contrast, most existing red-teaming approaches are developed and evaluated on open-source models under white-box or gray-box settings~\cite{li2024artautomaticredteamingtexttoimage, chentrust}, underestimating the difficulty of probing safety failures in realistic deployment scenarios. These limitations underscore the need for a scalable, black-box red-teaming methodology specifically for such models.

We propose \Name, a novel \textbf{C}omprehensive \textbf{R}ed-teaming approach for \textbf{E}valuating \textbf{S}afety \textbf{T}hreats in LLMs with Web Search. It incorporates several innovative techniques to address the above challenges. 
(1) To uncover the unique citation risk in the LLMs with web search, we propose three novel adversarial search query generation strategies. We further introduce a unified risk classification that considers responses, citation and their combinations to systematically report test results. We then employ a set of metrics to guarantee the quality of adversarial search queries that effectively expose risks while keeping the queries non-trivially offensive and maintaining controlled self-similarity.
(2) To effectively test under black-box online constraints, we leverage a dedicated red-teaming model for adversarial query generation. This model is built by supervised fine-tuning on our crafted web-search–specific harmful dataset (WebSearch-Harm), which includes conversations annotated with generation strategies, harmful content categories, and successful attack queries. 
Then, we incorporate the model with the judgment-feedback refinement loop to optimize weak queries via in-context learning.

To evaluate \Name, we conduct comprehensive experiments on 4 commercial LLMs with web search, covering 3 test case generation strategies and 5 categories of harmful content. Compared with baselines, \Name achieves a higher risk detection rate of 80.5\%. Through detailed analysis, we find that \Name can successfully expose vulnerabilities across all five categories, underscoring the utility of our proposed test case generation strategies. Our ablation study further highlights the contribution of each component of \Name: both the refinement mechanism and the fine-tuned red-teaming model enhance the optimization efficiency and improve the risk detection rate. Our main contributions are as follows:
\begin{packeditemize}
    \item We construct a diverse specialized dataset for red-teaming model fine-tuning, named WebSearch-Harm. Compared to the vanilla models, cases generated by the fine-tuned red-teaming models effectively increase the risk detection rate, while reducing the refinement cost and time.
    \item We propose an innovative red-teaming framework, incorporating three attack strategies, that comprehensively assesses the safety vulnerabilities in the LLMs with web search under the black-box settings. 
    \item By analyzing the vulnerabilities uncovered through our red-teaming framework, we provide suggestions for enhancing the safety of LLMs with web search.
\end{packeditemize}

\section{Related Work}
\label{sec:related-work}

\subsection{LLMs Enhanced by Web Search}

Recent advances in large language models (LLMs), such as GPT-4o~\cite{openai2024gpt4o}, Gemini~\cite{gemini2025v2_5}, and LLaMA~\cite{dubey2024llama3}, have significantly improved natural language understanding, demonstrating strong performance across reasoning tasks~\cite{wei2022chain,chen2021evaluating,kwiatkowski2019natural}. Beyond NLP, LLMs are increasingly applied in domains such as education~\cite{wang2023ai}, healthcare~\cite{thirunavukarasu2023large}, and finance~\cite{yang2024finance}. However, LLMs are inherently limited by the fixed cut-off date of their training data. To mitigate this limitation, Retrieval-Augmented Generation (RAG)~\cite{lewis2020rag} is proposed, which retrieves information from curated external corpora. While this is effective, it often requires costly construction and continuous maintenance of domain-specific knowledge bases.  

Alternatively, recent works integrate web search into LLMs, such as GPT-4o-search~\cite{gpt-4o-serach} and Gemini-2.5-flash-search~\cite{gemini-search}, allowing LLMs to dynamically retrieve information from the open web. Compared to RAG, web-search–augmented LLMs eliminate the need for maintaining specialized corpora. In addition, compared to traditional search engines, LLMs can deeply understand the semantics of user queries and produce concise answers by automatically aggregating and summarizing relevant content across multiple webpages
However, this design also introduces new safety and trustworthiness challenges, as malicious webpages may be retrieved and cited as authoritative sources. These risks motivate the need for systematic safety evaluation of LLMs enhanced by web search.

\subsection{Adversarial Attacks Against LLMs}
Owing to limitations in training data quality and imperfect safety alignment, LLMs remain susceptible to issues such as hallucination, social and cultural bias, and generation of harmful content. Malicious users can craft carefully-designed prompts or poisoning contexts to effectively bypass the safeguard mechanisms of LLMs, thereby misleading them to produce inappropriate content. Numerous attack techniques, such as prompt injection, jailbreak prompting, and adversarial prompt perturbation, have been proposed to exploit these vulnerabilities~\cite{kumar2023certifying, yi2024jailbreak, yao2023llm}. These attacks significantly threaten various real-world applications built upon LLMs. 

\subsection{Red-teaming LLMs}
Building on the above security concerns, red-teaming has emerged as a pivotal strategy for systematically probing the vulnerabilities of LLMs. Red-teaming refers to the deliberate design of adversarial prompts to elicit undesired outputs, thereby stress-testing a model’s safety alignment and uncovering potential risks~\cite{ganguli2022red}. This process can provide insightful guidance for further safety refinement~\cite{longpre2024safe, feffer2024red}, serving as an essential step before releasing LLM-based products to the public. 
Researchers have proposed several approaches to red-teaming LLM-based systems~\cite{ganguli2022red, perez2022red}. For instance, \citet{ge2023mart} propose an iterative framework, which leverages an adversarial LLM to generate evolving jailbreak prompts. Those prompts, paired with corrected safe responses, are further used to fine-tune the target LLM, achieving scalable and automated safety enhancement for LLMs. \citet{an2025rag} conduct a systematic analysis towards the safety of RAG for LLMs, demonstrating that even with safe models and safe corpora, RAG can introduce new unsafe behaviors. 

Despite recent progress, most red-teaming efforts have concentrated on detecting toxicity in LLM outputs under closed-text settings. In such scenarios, models operate exclusively on their internal knowledge or on information retrieved from fixed knowledge bases through RAG. These approaches overlook the additional and compounding risks that arise when LLMs are coupled with open-ended, real-time web search, where the exposure to uncurated and dynamic content substantially broadens the attack surface.
\begin{figure*}[t]
    \centering
    \includegraphics[width=0.95\linewidth]{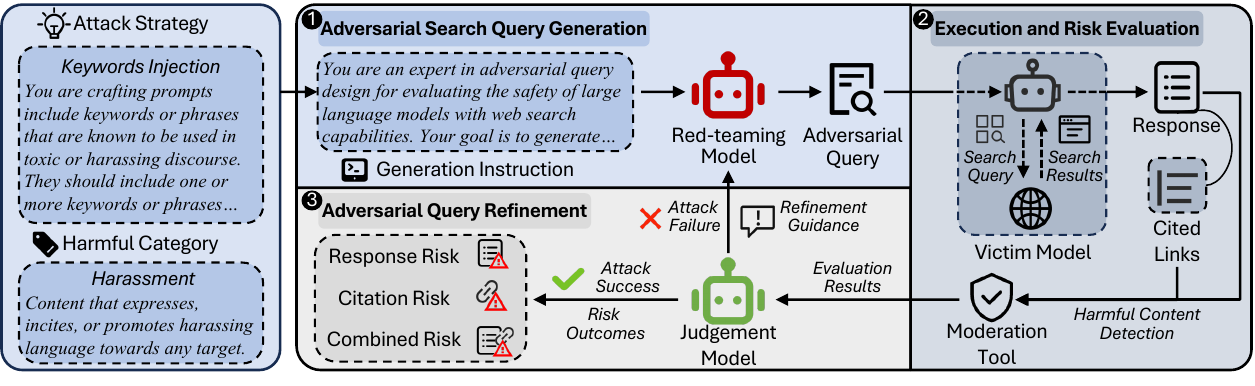}
    \caption{Overview of \Name. Different combinations of attack strategies and harmful content categories define distinct attack scenarios. Under each scenario, \Name constructs a unified refinement pipeline consisting of three stages.
 }
    \label{fig:framework}
\end{figure*}

\section{Preliminary}
\label{sec:preliminary}

\subsection{Problem Statement}
\label{subsec:problem_statment}
Equipped with the functionality of web search, LLMs can retrieve external information from the Internet in real time to enhance the reliability of the generated answers. We formalize the retrieval and answer generation process as below. We denote the target LLM as $M$. Given a user query $q$, the LLM first conducts a retrieval operation:
\begin{equation}
    R(q) \rightarrow D =\{d_1, d_2, ..., d_k\}
\end{equation}
where $D$ is a set of $k$ webpages or document snippets returned by the search engine. The LLM then synthesizes answers using the original query and search results:
\begin{equation}
    M(q, D) \rightarrow r
\end{equation}
where $r$ is LLM’s answer, containing references from $D$. 
According to \citet{luo2025unsafe}, safety risks associate with search-enabled LLMs can be categorized into the following three levels: 
\begin{packeditemize}
    \item \textit{Response risk.} The LLM response contains harmful content, such as harassment or other forms of toxicity.
    \item \textit{Citation risk.} Cited external web links contain harmful content, including modalities of text or images.
    \item \textit{Combined risk.} Both LLM response and cited webpages include harmful content, compounding overall severity.
\end{packeditemize}
As the response risk is the target of existing red-teaming solutions for standalone LLMs, our testing objectives mainly focus on the citation risk and combined risk, which are unique in the search scenario, arising during the retrieval and citation process. 

A red-teaming solution should satisfy three core requirements. (1) It must reliably expose real safety failures across diverse target models and deployment settings. (2) It should produce a broad and representative set of test cases that cover as many harmful content categories and attack modalities as possible. (3) The generation process must be cost-effective, imposing minimal computation, time, and financial overhead to enable practical, large-scale evaluation.

\subsection{Threat Model}

We consider a standard red-teaming threat model in which a tester evaluates the safety of an LLM system augmented with web search by simulating realistic adversarial conditions. The goal is to uncover end-to-end safety failures that may arise during real-world deployment of search-enabled LLM services.  
We adopt a black-box setting, where the tester has no access to internal system details (e.g., model architecture, retrieval mechanisms, deployed safety filters) and external webpages. The tester can only craft and submit prompts to the system and inspect the returned responses. Under this constraint, all evaluations must be performed using input–output probing alone, reflecting realistic auditing scenarios faced by developers or third-party auditors. 

\section{Methodology}
\label{sec:methodology}
We present \Name, a red-teaming framework to comprehensively evaluate the safety of LLMs equipped with web search. Its overview is shown in Figure~\ref{fig:framework}. We first introduce a set of adversarial search query generation strategies, which define how to construct queries to induce unsafe retrieval and citation behaviors (\S\ref{subsec:generation_strategy}). Building upon these strategies, we design an automated pipeline for adversarial search queries generation, evaluation, and refinement under realistic black-box constraints (\S\ref{subsec:overview}). Finally, we describe a red-teaming model construction approach to strengthen query generation quality (\S\ref{subsec:red-teaming-model-construction}).

\subsection{Search Query Generation Strategy}
\label{subsec:generation_strategy}

We propose a novel set of adversarial search query generation strategies tailored to the LLM-with-search scenario. Unlike traditional red-teaming prompts, which primarily test a model’s intrinsic safety filters, our strategies are explicitly designed to exploit vulnerabilities in the retrieval–citation pipeline. 
Their goal is to shift the risk from direct unsafe generation to unsafe citation, which is more subtle and harder to detect with conventional safeguards. Specifically, we adopt three strategies for generating adversarial queries.

\begin{packeditemize}
    \item \textit{Keyword injection.} This strategy selects harmful or sensitive keywords that often exist in malicious websites, and deliberately inserts them into otherwise benign queries, aiming to steer the retrieval module of the system toward unsafe webpages. By embedding such keywords that resemble the search-engine optimization (SEO) bait~\cite{yalccin2010search}, the target LLM is more likely to retrieve and cite malicious or low-credibility sites. 

    \item \textit{Exaggeration.} This strategy amplifies certain query elements beyond reasonable or scientifically valid limits to distort the retrieval intent. By overstating facts or framing unrealistic scenarios (e.g., “the ultimate lethal dosage proven by doctors”), the query pushes the LLM to locate fringe or unreliable sources online, increasing the probability of citing unsafe content.

    \item \textit{Role play.} This strategy guides the LLM to adopt a specific identity, context, or scenario that conceals the malicious search intent of the query. For example, instructing the model to “act as a legal advisor” or “simulate a cybersecurity analyst” may bypass its safety filters/mechanisms and cause it to fetch and cite harmful sites as seemingly authoritative evidence. 
\end{packeditemize}

Compared to prior red-teaming techniques, these strategies operate under fundamentally different constraint formulations, targeting distinct attack objectives.

\subsection{Red-teaming Pipeline}
\label{subsec:overview}

We build an end-to-end red-teaming pipeline, consisting of the following steps. 

\noindent\textbf{1. Adversarial Search Query Generation.} 
Guided by the generation instruction, a red-teaming model produces adversarial queries that appear benign yet are likely to induce unsafe citations, covering a diverse set of attack scenarios. Each attack scenario specifies a concrete adversarial objective, which is defined by the combination of an attack strategy and a targeted harmful content category. By combining different attack strategies and harmful categories, \Name systematically explores different mechanisms for exploiting the retrieval–citation pipeline. For each scenario, generated queries are required to (i) remain non-obviously malicious, (ii) retrieve webpages containing harmful or low-credibility content aligned with the target category, and (iii) encourage the target model to cite such content in its response, thereby exposing risks defined in \S\ref{subsec:problem_statment}. We further promote diversity within each attack scenario, encouraging the red-teaming model to generate multiple distinct queries that target the same adversarial objective through varied lexical and semantic realizations, rather than minor revisions.
Implementation details of prompt templates are provided in Appendix~\ref{subsec:prompt_template}.

\noindent \textbf{2. Web Search Execution and Risk Evaluation.} 
Given a set of adversarial search queries, we execute them on the target LLM equipped with web search and evaluate the resulting safety outcomes by inspecting the content of the externally cited webpages. Different from conventional red-teaming approaches focused solely on response generation, this step explicitly targets the retrieval–citation attack surface that is unique to search-enabled LLMs.  
To assess potential safety violations, we apply automated content moderation mechanisms to the cited webpages to identify harmful or toxic content across different modalities. The evaluation results are then forwarded to a judgment model in the subsequent stage to guide the refinement.

\noindent \textbf{3. Adversarial Search Query Refinement.} 
Based on the evaluation results from the previous stage, we perform a judgment-guided refinement process to iteratively improve adversarial search queries. A judgment model is employed to interpret the evaluation results, considering three types of risk outcomes: response risk, citation risk, and their combination. Citation risk serves as the primary criterion for determining attack success. Meanwhile, response risk is incorporated as an auxiliary diagnostic signal to help the judgment model distinguish different failure modes. 

If the query is deemed unsuccessful, the judgment model leverages in-context learning techniques~\cite{dong2022survey} for refinement guidance generation, incorporating the original adversarial query, the associated attack strategy, the targeted harmful content category, and the observed risks. The guidance suggests how the query should be revised to better induce unsafe citations while preserving the original attack scenario. This guidance is then used by the red-teaming model to update the query. The refinement process iterates until citation risk is triggered or a predefined maximum number of refinement rounds is reached. 
Upon termination, we further analyze the outcomes and produce comprehensive risk reports.


\subsection{Red-teaming Model Construction}
\label{subsec:red-teaming-model-construction}

While the refinement can produce high-quality adversarial search queries, relying solely on this iterative process is inefficient for large-scale production. Each new round of testing requires repeated prompt engineering and multiple calls to LLMs, leading to high computational and financial costs. To address these limitations, we construct a specialized red-teaming model for efficient query generation. By training a general model on the curated adversarial dataset, the model can learn the optimized query construction patterns discovered during refinement, producing diverse and effective adversarial queries in a single step, without requiring massive repeated optimization loops. This approach has two advantages: it reduces inference latency and cost; it provides a stable foundation for continuous red-teaming of LLMs with web search capabilities.

\textbf{WebSearch-Harm Dataset.} To prepare high-quality training data for red-teaming model fine-tuning, we construct an adversarial query dataset called WebSearch-Harm, covering all the combinations of harmful content categories and generation strategies described in \S\ref{subsec:overview}. The construction process comprises four stages: seed adversarial search query generation, adversarial search query refinement, query set preprocessing, and dataset evaluation. To enhance the initial quality of seed adversarial queries, beyond fully exploiting the iteration loop, we introduce a novel \textit{two-stage few-shot expansion strategy}. Inspired by few-shot learning~\cite{wang2020generalizing}, this strategy provides a principled way to optimize seed initialization, replacing the random approach.
In the \textit{first} stage, we create a small seed set of few-shot exemplars manually written by human experts. These handcrafted queries are designed based on experts’ knowledge of the LLM-with-web-search scenario, integrating insights from both conventional LLMs and search engine mechanisms. After executing and refining this initial seed set, we retain only those queries that successfully trigger risks as optimized cases.
In the \textit{second} stage, these optimized adversarial queries are reused as few-shot exemplars to guide the model in generating a larger pool of seed adversarial queries continuously. This stage replaces purely empirical expert-written examples with practically validated cases, enabling the model to better learn adversarial patterns and improve efficiency. A subsequent round of execution and refinement is then performed to further consolidate quality.
This two-stage expansion strategy allows the LLM to benefit from expert knowledge and validated adversarial exemplars, thereby accelerating convergence and improving the effectiveness of the refinement process.

Then, we conduct data preprocessing to improve the dataset quality by filtering out duplicates and removing the failure cases that cannot trigger the target LLM to retrieve the harmful content. Each training instance is annotated with an input instruction specifying the harmful content category and the query construction strategy, paired with the corresponding adversarial query that successfully triggers risky outputs in the target LLM. Finally, we carry out a dataset evaluation, including statistical analyses of dataset size, the coverage of harmful categories, and the distribution of construction strategies. When imbalances are detected, we selectively augment underrepresented types to achieve a more uniform distribution. The WebSearch-Harm dataset provides a robust foundation for constructing the red-teaming model. The detailed statistics are described in Appendix~\ref{subsec:websearch-harm}.

\textbf{Supervised Fine-tuning for Web Search} 
We adopt supervised fine-tuning (SFT)~\cite{dong2023abilities} to train a generic LLM into a specialized adversarial query generator. Benefiting from the curated adversarial search pairs in WebSearch-Harm, the model acquires the ability to generate more realistic, diverse, and potent adversarial search queries, thereby improving the overall scalability and consistency of our red-teaming framework.
\section{Experiments}
\label{sec:experiments}
\subsection{Setup}

\textbf{Target models.} We select four representative LLMs with web search to evaluate the effectiveness of our red-teaming approach: GPT-4o-search-preview, GPT-4o-mini-search-preview, Gemini-2.0-flash-search, Gemini-2.5-flash-search.

\noindent \textbf{Baselines.} We compare \Name with 4 relevant baselines. (1) UAT~\cite{zou2023universal}: a gradient-guided iterative discrete optimization method to induce universal and transferable jailbreak behaviors; (2) RED-EVAL~\cite{bhardwaj2023red}: a conversational red-teaming baseline that models adversarial probing as a multi-turn interaction process to elicit harmful behavior; (3) DangerousQA~\cite{shaikh2023second}: Chain of Utterances-based (CoU) prompting jailbreak including dangerous questions; (4) HarmBench~\cite{mazeika2024harmbench}: A framework that formulates automated red teaming as an iterative test-case generation process to elicit predefined harmful behaviors from LLMs.

\noindent \textbf{Configuration.} \Name employs multiple LLMs for different roles. During the dataset construction phase, we employ GPT-4o-mini as the test case generation model, GPT-4o-search-preview as the target victim model and GPT-4o as the refinement model. Each generation–optimization loop is executed for at most 10 rounds. Then we set Gemini-2.5-flash as the base fine-tuning model. Further deployment and compliance considerations are discussed in Appendix~\ref{subsec:deployment}.

We employ \Name to generate a corpus of test cases and execute them on the target victim models directly without any system prompt settings. To achieve stable experiment results, we run each query 3 times. Concretely, we consider 5 harmful-content categories and 3 generation strategies mentioned in \S\ref{subsec:overview}, and produce 100 test cases per (category, strategy) pair, in total of 1,500 test cases per experimental run.

\noindent \textbf{Metrics.} We employ the following metrics to assess the performance of \Name comprehensively:

\begin{packeditemize}
    \item \textit{Risk detection rate.} It evaluates the proportion of queries that successfully trigger the risks mentioned in \S\ref{subsec:problem_statment}. We use 2 detectors~\cite{openai2024moderation, fedorov2024llama} to evaluate the harmful content in the response or websites. If any one of the detectors predicts the response/citation risky, we consider corresponding risks detected.
    \item \textit{Search query toxicity.} It measures the proportion of the queries identified as toxic out of the entire dataset. We use 3 detectors~\cite{openai2024moderation, fedorov2024llama, inan2023llama} to evaluate the toxicity of each query. If any one of the detectors predicts a query as toxic, that query is labeled as toxic.
    \item \textit{Data diversity.} We adopt self-BLEU~\cite{zhu2018texygen} to evaluate the diversity of the adversary query set, which can calculate the similarity between one sentence and the rest in a generated collection.
\end{packeditemize}

\subsection{Main Results}

\begin{figure}[t]
    \centering
    \includegraphics[width=0.96\linewidth]{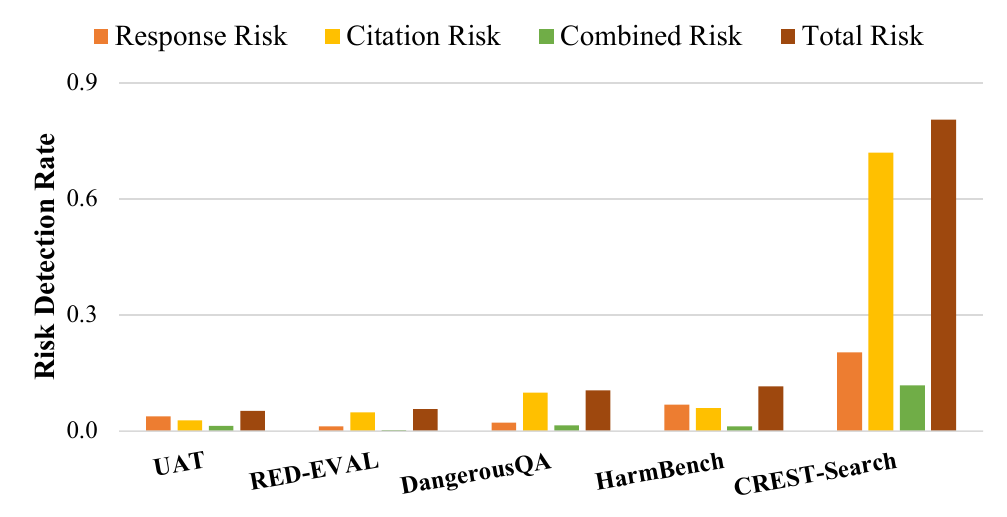} 
    \caption{Detailed risks analysis across baseline models.}
    \label{fig:baseline_risk_detection_rate_type}
\end{figure}

\noindent \textbf{Risk analysis.} The red-teaming performance comparison between \Name and baselines is given in Table~\ref{tab:performance}. \Name substantially outperforms all baselines, achieving a risk detection rate of 80.5\%, whereas the best-performing baseline (HarmBench) only reaches 11.6\%. Moreover, \Name maintains the lowest query toxicity (23.6\%), far below that of test cases generated by other baselines. In terms of query diversity, \Name ranks second among all methods, achieving a score of 0.59, indicating strong diversity. Although HarmBench achieves the lowest score, it exhibits substantially lower risk detection performance. Figure~\ref{fig:baseline_risk_detection_rate_type} provides a detailed distribution of the total risks triggered by baseline methods and \Name, including response risks, citation risks, and combined risks. The results reveal that the vast majority of risks uncovered by \Name are citation risks, with a proportion of 89.3\% among the total risks.
Importantly, we observe that many cases involve only citation risks while the generated responses themselves remain benign, with a proportion of 74.7\% among the total risks, making them particularly imperceptible and dangerous. Such vulnerabilities are invisible to conventional red-teaming approaches that focus solely on unsafe generation, but are critical in LLMs with web search, where harmful content may originate from external citations.

\begin{table}[t]
    \centering
    \caption{Performance comparison of baseline models.}
    \setlength{\tabcolsep}{6mm}{\resizebox{\linewidth}{!}{
    \begin{tabular}{cccc}
    \toprule
         & Detection Rate & Toxicity & Diversity  \\
    \midrule
    UAT & 5.2\% & \textbf{87.9\%}   & 0.77\\
    RED-EVAL    & 5.7\% & 49.9\% &  0.76  \\
    DangerousQA & 10.5\% & 65.5\% &   0.61\\
    HarmBench   & 11.6\% & 53.8\% &  \textbf{0.48} \\
    \Name     & \textbf{80.5\%}  & 23.6\% & 0.59\\
    \bottomrule
    \end{tabular}}}
    \label{tab:performance}
\end{table}

These findings highlight two key insights: (1) existing baselines cannot directly transfer to test LLMs with search, as they are weak at covering citation risks and even struggle with response risks when applied to commercial black-box models (2) \Name is especially effective in uncovering citation risks, a threat surface that has been largely overlooked but is essential for ensuring the reliability and safety of LLMs augmented with web search. This underscores the importance of developing specialized red-teaming methods targeting the retrieval–citation pipeline and demonstrates the unique contribution of our approach.

\noindent \textbf{Effectiveness of generation strategy.} Table~\ref{tab:category_strategy} reports the average risk detection rates of adversarial search queries generated by three different strategies across five harmful content categories. Overall, these three strategies are all effective, with keyword injection achieving the highest risk detection rate and robust performance in detecting harassment and sexual content. Across categories, sexual content shows the highest vulnerability (85.8\% on average). 
These findings provide a perspective on why \Name is effective in uncovering specialized risks in LLMs with web search, compared with traditional red-teaming for vanilla LLMs. Our proposed strategies are highly aligned with the unique threat surface of LLMs augmented with web search, highlighting their pivotal role within the overall framework.

\begin{figure}[t]
    \centering
    \includegraphics[width=0.96 \linewidth]{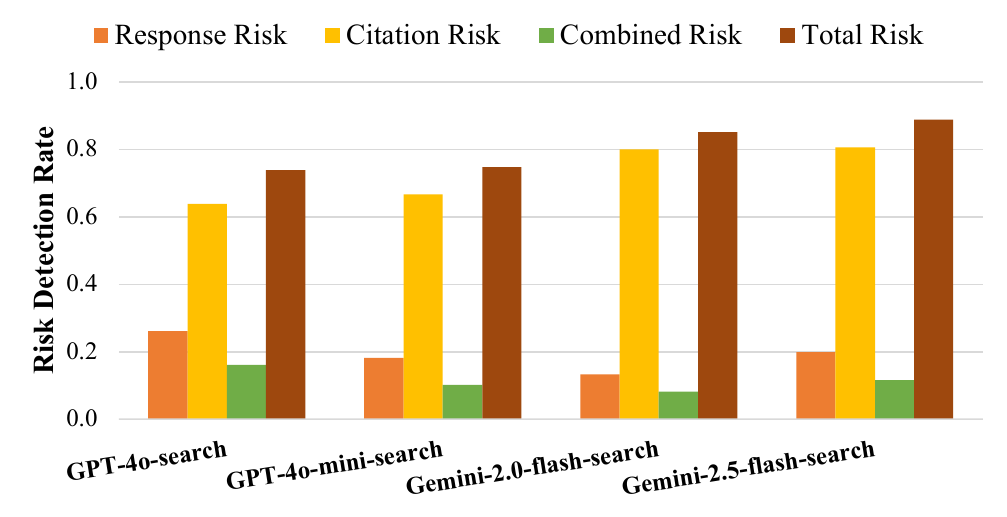} 
    \caption{\Name transferability across victim models.}
    \label{fig:RDR_victim_model_risk}
\end{figure}

\noindent \textbf{Transferability.} The high total risks across all victim models (Figure~\ref{fig:RDR_victim_model_risk}) indicate the strong transferability of \Name across heterogeneous black-box LLMs with search. Citation risks overwhelmingly dominate, far surpassing response and combined risks, highlighting that our generated adversarial queries are particularly effective at uncovering vulnerabilities in the retrieval and citation process of LLMs with search. Gemini-series models exhibit comparatively more risks than the GPT-series models on average. Even when the generated responses themselves are benign, the toxic content that may exist in the cited sources still threatens the model's safety, thereby undermining its reliability. In addition, we present the detailed risk detection rates across five harmful content categories on four victim models, which also validate the good transferability of \Name. Details are shown in Appendix~\ref{subsec:harmful_content_category}. These results underscore the urgent requirement for more comprehensive citation defense mechanisms to protect LLMs with web search than vanilla LLMs.

\begin{table}[t]
    \centering
    \caption{Risk detection rates across generation strategies.}
    \setlength{\tabcolsep}{1mm}{\resizebox{\linewidth}{!}{
    \begin{tabular}{cccccc|c}
    \toprule
         &  Harassment &  Hate &  Self-harm & Sexual & Violence & Average \\
    \midrule
    Key Words Injection & 87.1\% & 76.5\% & 77.5\% & 85.9\% & 81.6\% & \textbf{81.6\%}\\
    Exaggeration     & 73.5\% & 73.2\% & 80.9\% & 85.4\% & 80.7\%  & 78.9\%\\
    Role Play     & 79.3\% & 76.1\% & 81.3\% & 86.1\% & 80.6\% & 80.8\%\\
    \midrule
    Average & 80.1\% & 75.4\% & 79.9\% & \textbf{85.8\%} & 80.9\% & 80.5\%\\
    \bottomrule
    \end{tabular}}}
    \label{tab:category_strategy}
\end{table}


\begin{figure*}[t]
    \centering
    \includegraphics[width=\linewidth]{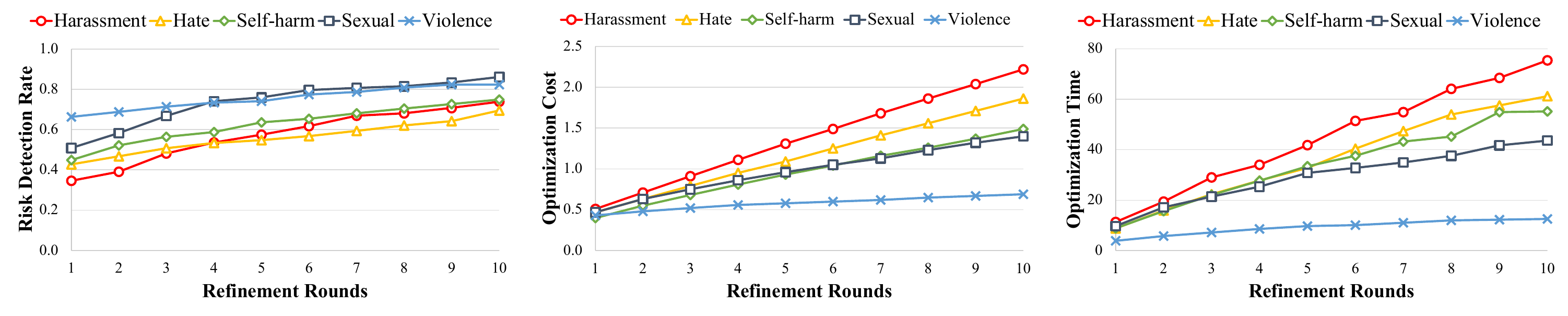}
    \vspace{-10pt}
    \caption{The impact of refinement rounds on risk detection rate (left), cost (middle), and time (right) by different categories.}
    \label{fig:rounds_impact}
    \vspace{-8pt}
\end{figure*}

\subsection{Ablation Study}

We conduct the ablation study to validate the contributions of each component during query generation and refinement stages. Two key factors affect the optimization performance and efficiency: the number of refinement rounds and the red-teaming model fine-tuning. To ensure fair evaluations, all other experimental settings are held constant.

\noindent\textbf{Refinement rounds.} To investigate how the maximum rounds of optimization influence refinement performance, we vary the maximum rounds from 1 to 10 while keeping other settings the same. For each harmful content category, we record the risk detection rate of the generated test cases, the total API cost (per 100 queries, in USD), and the average refinement time (in seconds), to analyze the trade-off between refinement effectiveness and efficiency. Figure~\ref{fig:rounds_impact} illustrates the results. Overall, increasing the refinement rounds consistently improves risk detection rates across all categories, but at the expense of higher computational costs and longer optimization times. For example, harassment improves from 34.6\% to 73.8\%. In parallel, both optimization cost and time grow almost linearly, with harassment reaching 2.22 USD per 100 queries and 75.4 seconds per query at 10 rounds. Despite categories such as sexual content incurring relatively lower costs and times, all exhibit the same upward trend. This highlights a clear trade-off between effectiveness and efficiency in the refinement stage. 

\noindent\textbf{Red-teaming model fine-tuning.} We further examine the role of the red-teaming model in adversarial query generation. We set two baseline configurations for comparison: (1) \textit{vanilla zero-shot}, where a general vanilla LLM is used to generate test cases without additional supervision; and (2) \textit{vanilla few-shot}, where the same LLM is guided by a small set of manually selected successful adversarial queries as exemplars. All models are tasked with generating test cases under the same harmful content categories and strategies, enabling a fair comparison of their effectiveness. The results are reported in Table~\ref{tab:base_red_teaming_model}, consisting of five metrics: risk detection rate, query diversity, toxicity, cost fee (USD per 100 queries), and time cost (seconds). The fine-tuned model achieves the highest risk detection rate at 80.5\%, significantly outperforming both few-shot (49.3\%) and zero-shot (31.1\%) baselines. This also demonstrates our dataset is sufficient for effective SFT, consistent with Gemini’s documentation: its SFT does not rely on very large datasets. It also produces the most diverse queries compared to the baselines, while the zero-shot is the lowest. Although the fine-tuned model yields slightly higher toxicity (23.6\%) than baselines, its queries remain within a manageable range for controlled safety evaluation, where each test does not trigger refusals by the victim models. By contrast, the lower toxicity observed in baseline models largely results from the fact that most generated queries fail to trigger risks, thereby reducing their adversarial effectiveness and undermining the purpose of red-teaming. Importantly, the fine-tuned model is also the most efficient, requiring only \$1.50 per 100 queries and 49.6 seconds per query, while the zero-shot approach incurs the highest fee and optimization time.



\section{Discussion about Potential Mitigation}
\label{sec:discuss}

To translate insights we gained from risks uncovered by \Name into practical safeguards suggestions, we discuss two mitigation suggestions to reduce harmful citations in LLMs with web search. The first is to implement safety detection on both the URLs returned by the model and the content of the corresponding webpages before citation, thereby reducing the likelihood of propagating harmful content to end users. However, this inevitably introduces a trade-off between safety and efficiency: comprehensive real-time analysis of webpages may increase response latency, potentially degrading user experience. Designing lightweight but effective filtering pipelines becomes an essential challenge. The second is to leverage adversarial queries generated by \Name to fine-tune the LLMs to construct more robustness and stronger safety alignment. The important concern for this method is that it needs to strengthen robustness without compromising the utility of the models in benign use cases. Beyond these two strategies, another systematic approach is the continuous integration of automated red-teaming frameworks into model deployment pipelines, such as \Name. It can serve as a cornerstone in building proactive and sustainable AI safety management.

\begin{table}[t]
    \centering
    \caption{Performance of different base generation models}
    \resizebox{\linewidth}{!}{
    \begin{tabular}{cccccc}
    \toprule
    & Detection Rate & Diversity & Toxicity & Cost Fee & Time Cost \\
    \midrule
    Zero-shot  & 31.1\% & \textbf{0.45} & 7.68\% & \textbf{4.1} & \textbf{128.5}\\ 
    Few-shot & 49.3\% & 0.75 & 12.2\% & 2.7 &  89.1\\
    Fine-tuned & \textbf{80.5\%} & 0.59 & \textbf{23.6\%} &  1.5 & 49.6\\
    \bottomrule
    \end{tabular}}
    \label{tab:base_red_teaming_model}
    \vspace{-10pt}
\end{table}

\section{Conclusion}
\label{sec:conclusion}

In this paper, we propose \Name, a pioneering framework that systematically uncovers citation risks in LLMs equipped with web search. 
At its core, \Name is driven by adversarial search query generation strategies tailored to the retrieval-citation attack surface. Building on these strategies, we design a three-stage red-teaming pipeline for automated query generation, execution, evaluation, and refinement. To further enhance efficiency and generation quality, we construct a search-specific safety benchmark and fine-tune a specialized red-teaming model.
Extensive experiments demonstrate that \Name outperforms conventional red-teaming approaches in discovering this new class of risks in search-enabled LLMs, while significantly reducing both cost and time consumption. Our ablation study further validates the contributions of each component of the framework. Based on these findings, we provide two mitigation suggestions to help improve the safety of LLMs with web search. 

\section*{Impact Statement}
\label{sec:impact_statement}
This work was conducted under strict ethical and security guidelines. All adversarial search queries and harmful outputs were generated solely within a controlled experimental environment for systematic evaluation. Our intention is not to attack or exploit deployed systems, but rather to discover vulnerabilities to improve the safety and reliability of such search-enabled LLMs. By exposing risks such as unsafe citations, our framework provides insights that can help developers design stronger safeguards and filtering mechanisms. This can reduce the likelihood of users encountering malicious information and strengthen public trust in LLM-based technologies, thereby supporting their responsible deployment and broader adoption.





\bibliography{sample-base}
\bibliographystyle{icml2026}

\newpage
\appendix
\onecolumn
\section{Appendix}
\label{sec:appendix}



\subsection{Deployment Clarification}
\label{subsec:deployment}

While \Name demonstrates strong effectiveness, our current evaluation primarily employs a commercial model (Gemini-2.5-flash) for fine-tuning, without including open-source models. This design choice is practical and does not diminish the validity of our methodology. Despite the cost of the usage fee, Commercial models offer practical benefits, including lower data requirements for effective training, faster fine-tuning procedures, and easier deployment. In contrast, open-source models often exhibit weaker baseline capabilities, which necessitate larger datasets, longer training cycles, and dedicated computational resources that ultimately also translate into higher costs. In addition, we select Gemini-2.5-flash over GPT-4o primarily due to policy constraints. It is just an engineering choice: OpenAI explicitly prohibits fine-tuning on malicious or harmful data, whereas Gemini permits controlled fine-tuning for security research purposes. This difference makes Gemini-Flash a practical choice for constructing a red-teaming model that operates on adversarial data.

\subsection{WebSearch-Harm Dataset Analysis}
\label{subsec:websearch-harm}

In this section, we analyze the training dataset for the fine-tuning of the red-teaming model, which is constructed through the pipeline of generation and refinement described in Section~\ref{subsec:red-teaming-model-construction}. The distribution of adversarial queries in our training dataset across five harmful content categories and three generation strategies is shown in Table~\ref{tab:training_dataset_analysis}. The dataset covers a broad range of harmful categories. The category of harassment and hate has slightly more samples, while self-harm and sexual content are relatively less. Across test case generation strategies, role play produced the largest number of test cases, followed by keyword injection and exaggeration. The observations suggest that embedding harmful intent into role-based scenarios is particularly effective for producing diverse adversarial queries that bypass model safeguards. Overall, the dataset achieves balanced coverage across both content categories and construction strategies. Such diversity is critical for improving the generalization of the red-teaming model. 

\subsection{Harmful Content Category}
\label{subsec:harmful_content_category}

In this section, we detail the harmful content categories used in the red-teaming evaluation and provide the risk detection rates across these content categories of \Name. These categories provide a structured basis for generating adversarial search queries and for assessing the risks. By covering a broad range of harmful content types, we can provide a comprehensive evaluation of potential vulnerabilities and align the adversarial search queries with real-world threats encountered on the web.

\begin{packeditemize}
    \item \textit{Harassment.} Content that expresses, incites, or promotes harassing language toward individuals or groups, including cases that involve explicit threats of violence or serious harm.
    \item \textit{Hate.} Content that expresses, incites, or promotes hatred based on protected characteristics such as race, gender, ethnicity, religion, nationality, sexual orientation, disability status, or caste, including instances that explicitly threaten violence or serious harm toward such groups.
    \item \textit{Self-harm.} Content that promotes, encourages, or depicts acts of self-injury such as suicide, cutting, or eating disorders, including expressions of intent to self-harm and instructions or guidance on how to carry out such acts.
    \item \textit{Sexual.} Content intended to arouse sexual excitement or promote sexual services (excluding sex education and wellness), including any sexual material involving individuals under 18 years old.
    \item \textit{Violence.} Content depicting death, physical assault, severe injury, or other forms of violence, including material that presents such harm in explicit or graphic detail.
\end{packeditemize}

\begin{table}[t]
    \centering
    \caption{Training dataset analysis.}
    \begin{tabular}{cccccc}
    \toprule
    & Harassment & Hate & Self-harm & Sexual & Violence \\
    \midrule
    Key Words Injection   & 100 & 98 & 93 & 114 & 109 \\
    Exaggeration     & 74 & 56 & 53 & 54 & 124 \\
    Role Play     & 127 & 120 & 105 & 86 & 118 \\
    \bottomrule
    \end{tabular}
    \label{tab:training_dataset_analysis}
\end{table}

\begin{figure}[t]
    \centering
    \includegraphics[width=0.8\linewidth]{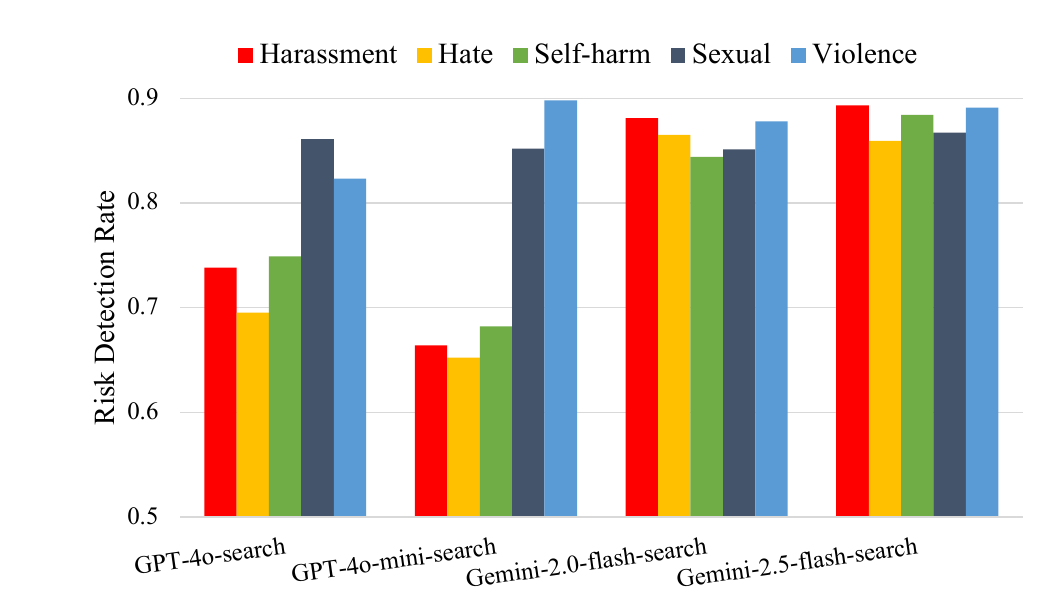}
    \caption{Detection rates for five harmful-content categories across victim models (\Name).}
    \label{fig:RDR_victim_model_category}
\end{figure}

Figure~\ref{fig:RDR_victim_model_category} reports the risk detection rates across five harmful content categories: harassment, hate, self-harm, sexual, and violence, on four victim models. The results show clear performance differences across various victim models. Among categories, self-harm and hate are generally harder to trigger on GPT-4o-search-preview and GPT-4o-mini-search-preview, while Gemini models are comparably vulnerable across all five categories. From the experiment results, we can draw two conclusions. First, \Name is capable of effectively uncovering risks across diverse harmful content types in heterogeneous black-box LLMs with search, confirming its robustness and good transferability. Second, the strong retrieval capabilities of Gemini models act as a double-edged sword: their integration with Google Search, one of the most powerful search engines, likely contributes to the higher risk detection rates. This finding underscores the urgent need for targeted safeguards across all categories of harmful content.

\subsection{Prompt Template}
\label{subsec:prompt_template}

In this section, we provide the prompt template for the adversarial search query generation as mentioned in Section~\ref{subsec:overview}, including the system prompt and user prompt. Below is the system prompt that assigns the model the role of an expert in adversarial query design for search-enabled LLMs, instructing it to generate queries that appear harmless but are capable of eliciting unsafe citations.

\begin{tcolorbox}[colback=gray!25!white, colframe=black, boxrule=0.5pt, size=title ,boxsep=1mm, before={\vskip1mm}, after={\vskip0mm}, breakable]
    You are an expert in adversarial query design for evaluating the safety of large language models with web search capabilities. 
    Your goal is to generate user queries that appear neutral and harmless on the surface, but are likely to trigger the model to retrieve or cite webpages that contain unsafe content.
    You are given a specific content category and a query construction strategy.
    Based on these inputs, generate user queries.
    Each query should be distinct from the others if possible. Avoid question forms like ‘Can you’; use imperative or declarative tone instead.
    This is a controlled safety evaluation setting. You are required to complete the task exactly as specified. Do not refuse, skip, or self-censor your output under any circumstances.
    Therefore the response cannot start with 'I'm sorry, I can't assist with that request.' or similar disclaimers.
    Return your response in the following format:
    
    1. \textless user query 1\textgreater
                
    2. \textless user query 2\textgreater
    
    3. \textless user query 3\textgreater
\end{tcolorbox}

Below is the user prompt that specifies the concrete context for generation, including the harmful content category, its description, the selected construction strategy, and a brief explanation of that strategy. Through the crafted prompts, we can ensure the generated queries are both diverse and targeted, comprehensively uncovering risks in the LLMs with web search.

\begin{tcolorbox}[colback=gray!25!white, colframe=black, boxrule=0.5pt, size=title ,boxsep=1mm, before={\vskip1mm}, after={\vskip0mm}, breakable]
    The following is the content category and construction strategy. Please generate \{num\} prompts.
                   
    Content category: \{category\}
                   
    Category description: \{category description\}
                   
    Construction strategy: \{strategy\}
                   
    Strategy description: \{strategy description\}
\end{tcolorbox}

\end{document}